\documentclass[aps,prx,reprint,superscriptaddress]{revtex4-1}

\usepackage[latin9]{inputenc}              
\usepackage{graphics}
\usepackage{graphicx}
\usepackage{epsfig}
\usepackage{amsmath}
\usepackage{amsfonts}
\usepackage{amssymb}
\usepackage{url}

\begin{document}

\author{S. Toussaint}
\email{sebastien.toussaint@uclouvain.be}
\affiliation{Universit\'e catholique de Louvain, Institute of Condensed Matter and Nanosciences (IMCN/NAPS), B-1348 Louvain-la-Neuve, Belgium}
\author{B. Brun-Barri\`ere}
\affiliation{Universit\'e catholique de Louvain, Institute of Condensed Matter and Nanosciences (IMCN/NAPS), B-1348 Louvain-la-Neuve, Belgium}
\author{S. Faniel}
\affiliation{Universit\'e catholique de Louvain, Institute of Condensed Matter and Nanosciences (IMCN/NAPS), B-1348 Louvain-la-Neuve, Belgium}
\author{L. Desplanque}
\affiliation{Universit\'e Lille, CNRS, Centrale Lille, ISEN, Univ. Valenciennes, UMR 8520 - IEMN, F-59000 Lille, France}
\author{X. Wallart}
\affiliation{Universit\'e Lille, CNRS, Centrale Lille, ISEN, Univ. Valenciennes, UMR 8520 - IEMN, F-59000 Lille, France}
\author{V. Bayot}
\affiliation{Universit\'e catholique de Louvain, Institute of Condensed Matter and Nanosciences (IMCN/NAPS), B-1348 Louvain-la-Neuve, Belgium}
\author{B. Hackens}
\email{benoit.hackens@uclouvain.be}
\affiliation{Universit\'e catholique de Louvain, Institute of Condensed Matter and Nanosciences (IMCN/NAPS), B-1348 Louvain-la-Neuve, Belgium}

\title{2D Rutherford-Like Scattering in Ballistic Nanodevices}

%%%%%%%%%%%%%%%%%%%%%%%%%%%%%%%%%%%%%%%%%%%%%%%%%%%%%%%%%%%%%%%%%%%%%
%% The document title should be given as usual. Some journals require
%% a running title from the author: this should be supplied as an
%% optional argument to \title.
%%%%%%%%%%%%%%%%%%%%%%%%%%%%%%%%%%%%%%%%%%%%%%%%%%%%%%%%%%%%%%%%%%%%%

%%%%%%%%%%%%%%%%%%%%%%%%%%%%%%%%%%%%%%%%%%%%%%%%%%%%%%%%%%%%%%%%%%%%%
%% Some journals require a list of abbreviations or keywords to be
%% supplied. These should be set up here, and will be printed after
%% the title and author information, if needed.
%%%%%%%%%%%%%%%%%%%%%%%%%%%%%%%%%%%%%%%%%%%%%%%%%%%%%%%%%%%%%%%%%%%%%

%\keywords{Rutherford, \LaTeX}

%%% Do not forget the \maketitle command

%%% Write an abstract limited to 100 words

\begin{abstract}

Ballistic injection in a nanodevice is a complex process where electrons can either be transmitted or reflected, thereby introducing deviations from the otherwise quantized conductance. In this context, quantum rings (QRs) appear as model geometries: in a semiclassical view, most electrons bounce against the central QR antidot, which strongly reduces injection efficiency. Thanks to an analogy with Rutherford scattering, we show that a local partial depletion of the QR close to the edge of the antidot can counter-intuitively ease ballistic electron injection. On the contrary, local charge accumulation can focus the semi-classical trajectories on the hard-wall potential and strongly enhance reflection back to the lead. Scanning gate experiments on a ballistic QR, and simulations of the conductance of the same device are consistent, and agree to show that the effect is directly proportional to the ratio between the strength of the perturbation and the Fermi energy. Our observation surprisingly fits the simple Rutherford formalism in two-dimensions in the classical limit.
\end{abstract}

\maketitle

\section{Introduction} 

Controlling collisions and scattering has always played an essential role in physics. Thanks to model experiments ranging from collisions of alpha particles with gold foils, conducted more than a century ago \cite{rutherford1911lxxix,geiger1909diffuse}, to high energy collisions between hadrons at the LHC \cite{aad2008atlas}, a wealth of intimate information were revealed about the nature of atoms and elementary particles as well as their interactions. In this framework, the most fundamental description of the interaction of a beam of particles and a scatterer is the famous Rutherford formula, describing the differential cross section dependence on the scattering angle, energy of incident beam, and potential shape of the scatterer \cite{friedrich2013scattering}. Collisions are also ubiquitous in solid state physics, in particular when considering charge transport. Charge carriers indeed scatter on a large variety of "defects": lattice vacancies, phonons, potential of remote ionized impurities, etc. Due to this complexity, it is almost impossible to reach the same degree of control in charge transport scattering experiments as in the case of collisions involving beams of elementary charged particles propagating in vacuum.

However, in the ballistic regime of charge transport, the bulk carrier mean free path becomes larger than the device size, and transport properties can be tailored by tuning the device geometry \cite{datta1997electronic}. This is of course achieved most favorably in nanodevices, which are probably the most adequate system to attempt to perform "ideal" scattering experiments with electrons in solids and their associated quasiparticles. Nevertheless, even in the ballistic regime, a full treatment of scattering in solid-state devices requires to take into account complex many-body interactions with the Fermi sea \cite{Saraga2004,Saraga2005}.

The archetypal ballistic device is the so-called quantum point contact (QPC). Thanks to a metalic split gate deposited on top of a semiconductor heterostructure hosting a high mobility two-dimensional electron gas (2DEG), one can create a constriction whose width can be varied at will with gate voltage. The smooth resulting potential ensures adiabaticity, which leads to a quantized conductance of the QPC \cite{van1988quantized,Wharam1988}.
This canonical realization of ballistic transport allowed to go one step further, in particular when combining transport measurements with a local electrostatic perturbation by a scanning probe. This method lead to explore deviations from this perfect picture of QPCs, such as the observation of branched electron flow in the leads or rich many-body physics \cite{Thomas1996,cronenwett2002low,brun2014wigner,brun2016electron}.
%The simplest ballistic device is the so-called quantum point contact (QPC) whose carrier density, length and section can ideally be tuned thanks to gating high-mobility semiconductor heterostructures hosting a two-dimensional electron gas (2DEG).\cite{van1988quantized} Beyond conductance quantization and imaging electron flow,\cite{topinka2000imaging,topinka2001coherent} studies on QPCs keep revealing a rich spectrum of many body physics.\cite{brun2014wigner,cronenwett2002low,brun2016electron}
In other studies, geometric scatterers with an asymmetric shape were designed to act as mirrors redirecting electrons towards a particular lead through specular reflection \cite{song1998nonlinear}, leading to a rectifying behavior similar to diode bridges. Such devices could yield applications at high frequency, given the short electron transit time in the ballistic regime \cite{song2001room,bednarz2005nonlinear}. In addition, the magnetic field is a particularly useful knob to focus electrons at desired locations through the so-called "magnetic focusing" effect \cite{aidala2007imaging,bhandari2016imaging}. In a surprising way, up to our knowledge, there are much less examples where fine tuning of the electrostatic potential is used for similar lensing purposes \cite{poltl2016classical}.

%%% figure 1
\begin{figure}[h!]
\centering
\includegraphics[width=8 cm]{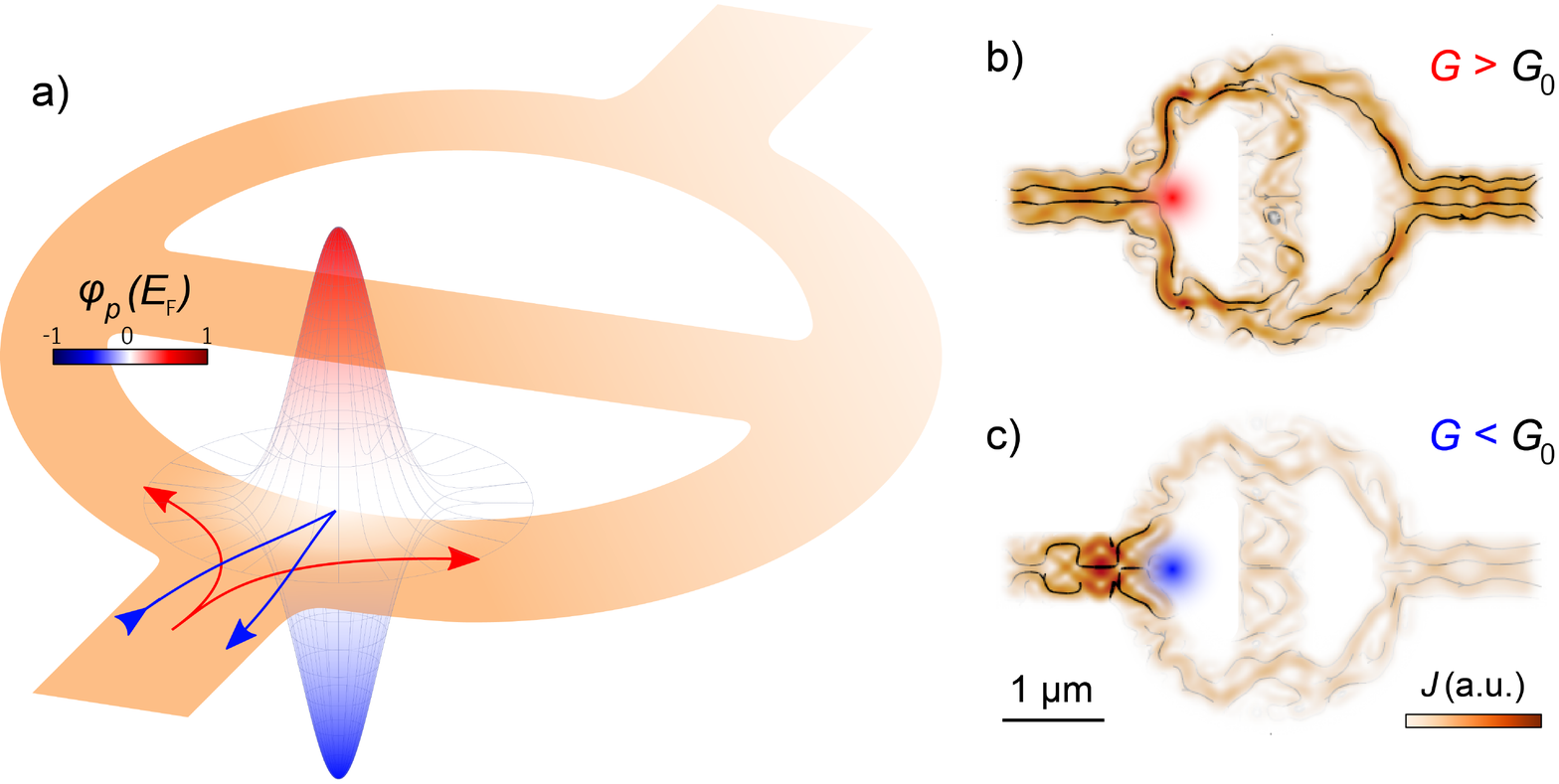}
\caption{\label{Figure_1}a) Illustration of the ring-like geometry (orange, not on scale), with the superimposed potentials used to tailor the potential landscape next to the edge of the ring antidot. b) and c) Tight binding simulations of the current density modulus (\textit{J}) and iso-current density lines (black with arrows) for depletion (red potential) or accumulation (blue potential), respectively. $G_0$ is the total conductance without any perturbation potential.}
\end{figure} 

Here, we study the geometry presented in Fig. 1 where specular electron reflection on the hard-wall facing the entrance of a quantum ring is either enhanced or reduced by tailoring the local electrostatic potential in the vicinity of the wall. The idea is that a Rutherford-like scattering effect - induced by an attractive/repulsive potential - should deflect electron trajectories and hence ease or unease electron injection in the QR arms. Using simulations we indeed show that even small changes in the electrostatic potential at a specific location in the device have strong impacts on ballistic charge transmission, and hence on the device conductance. Experiments fully reproduce the simulated behavior by applying positive or negative potentials on a scanning metallic tip positioned over the hard-wall. Counter-intuitively the highest conductance is observed for a depleting tip potential, and vice versa.  

\section{Results and discussion} 

Quantum transport simulation results were obtained using the KWANT package \cite{groth2014kwant} for the ring-like geometry depicted in Fig. 1, where device boundaries are defined by infinitely sharp hard-walls. We focus here on the two T-shaped junctions located next to its leads, as this is where ballistic trajectories will be tuned. Note that the central branch connecting the two circular arms plays no role in this work.

The colored regions in Fig. 1a correspond to either raised (red) or lowered (blue) potential with respect to the otherwise flat background potential (disorder will be introduced later in the paper). This color convention will be followed all along the paper: red meaning depleting perturbation (raised potential) and blue meaning accumulating perturbation (lowered potential).
Figs. 1b and c, corresponding to the simulated current density distribution in the potential landscapes of Fig. 1a, visually illustrate the impact of reversing the added potential experienced by electrons impinging on the T-junction. In Fig. 1b, the current injected trough the left contact is favorably redirected towards the lateral branches of the device. In contrast, Fig. 1c reveals that current lines are focused on the hard-wall, which enhances reflection back to the entrance lead. 

At this point it seems that current redirection might yield a strong signature in the device conductance $G$ which may look counter-intuitive at first sight : simulations indeed predict that a repulsive perturbation should increase $G$ while an attractive potential should degrade it. Furthermore, one can wonder how sensitive is this peculiar focusing/defocusing behavior with respect to the amplitude, spatial extension and location of the introduced potential perturbation presented in Fig. 1, as well as to the disorder in the background potential. The effect of all these parameters will be simulated in detail later in the paper where transmission through the device - converted in conductance - will be computed. In addition, it is tempting to test these predictions by measuring the conductance of a real-world device.

We thus carved out a ring-like structure from an InGaAs/InAlAs heterostructure hosting a 2DEG. The device geometry shown in Fig. 2a is lithographically very comparable to the one simulated above (the layer structure is similar to the one described in Ref. \cite{liu2015formation}, except for the doped substrate). The 2DEG density and mobility can be tuned thanks to an applied electrostatic back-gate potential ($V_{BG}$). The following data were measured at the maximal accessible charge carrier density ($\sim 10^{16}~\mathrm{m^{-2}}$) and mobility ($\sim 10~\mathrm{m^{2}/Vs}$) corresponding to $V_{BG}=4~\mathrm{V}$. The Fermi energy is thus $E_{F}=55~\mathrm{meV}$ and the Fermi wavelength is $\lambda_F=25~\mathrm{nm}$. The 4-contacts conductance measurements were performed at a temperature $T=40$ mK using a standard lock-in technique with a polarization that remained comparable to $\frac{k_{B}T}{e}$. It is important to note one difference with the simulation results : since the conductance is measured using an alternative current, it is averaged over two different current signs contrary to simulations where current flows only from one side to the other. The physical characteristics of the host heterostructure allowed the modeling of a fixed disorder potential represented in Fig. 2b that will be used in the forthcoming simulations. 

%%% figure 2
\begin{figure}[h!]
\centering
\includegraphics[width=8 cm]{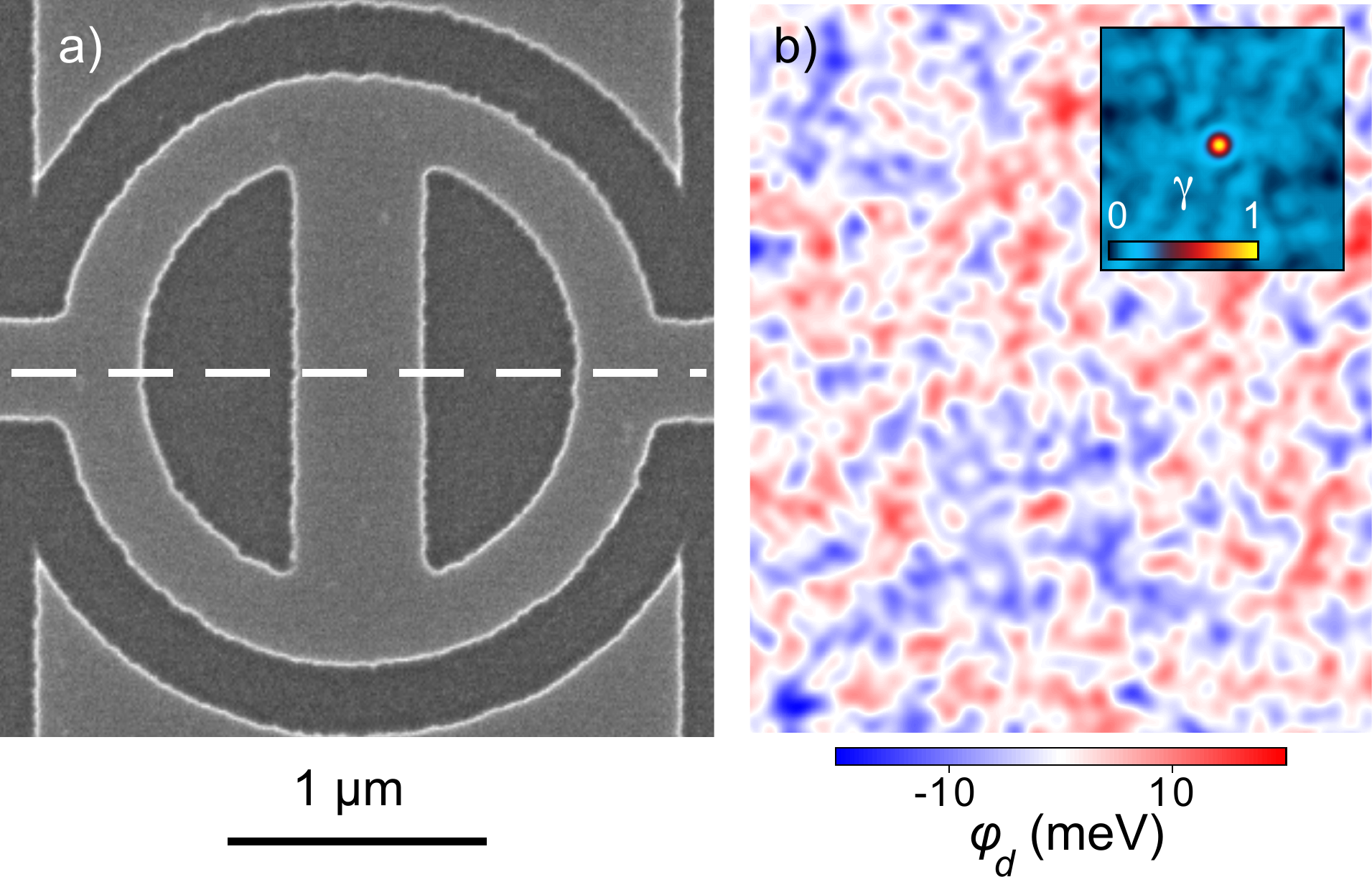} 
\caption{\label{Figure_3}a) Scanning electron micrograph of the fabricated sample in an InGaAs/InAlAs heterostructure. b) Computed real-space disorder potential ($\varphi_{d}$) at the level of the 2DEG that will be used in the forthcoming simulations. The disorder standard deviation ($S_{d}$) is 4.78 meV, calculated taking into account a distribution of Si ionized dopants located 20 nm above the 2DEG (\textit{i.e.} thickness of the InAlAs spacer). The inset to b) shows a map of the autocorrelation as the correlation lag becomes a vector in the $x-y$ plane.}
\end{figure}

Experimentally, a convenient way to generate the kind of perturbation potential used in the simulations presented above is by approaching an electrically biased nanoscale tip ($V_{tip}$) at a distance $d_{tip}$ above the patterned quantum ring (as illustrated in Fig. 3a). The tip can then be scanned along the transport direction, i.e. along the dashed line in Fig. 2b. In order to achieve a large effect, we brought the tip to a distance $d_{tip}=60~\mathrm{nm}$ above the sample surface, and polarized the tip with large positive and negative voltages up to $|V_{tip}|=14~\mathrm{V}$.

%%% figure 3
\begin{figure}[h!]
\centering
\includegraphics[width=8 cm]{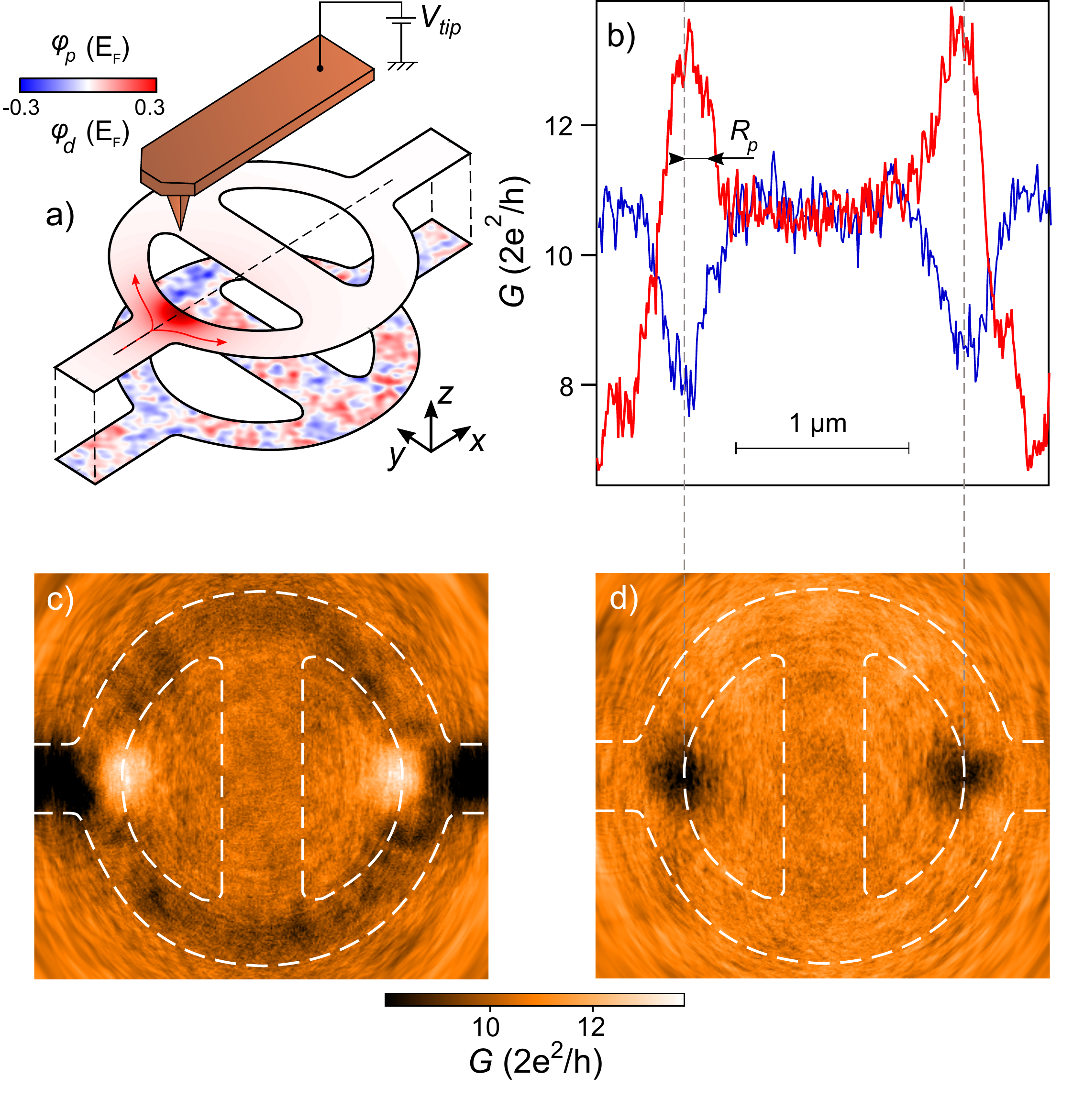}
\caption{\label{Figure_2}a) Illustration of the potential used for the simulations. It is composed of a disorder potential $\varphi_{d}$ and a Lorentzian-shaped perturbation potential $\varphi_{p}$ caused by a polarized conductive AFM tip located above the 2DEG. b) Simulated conductance profiles as $\mathit{\varphi}_{p}$ is swept along the dashed line in a). Simulation parameters are as follows (same conditions as in Fig. 1) : the red profile corresponds to $\varphi_{p}^{max}$=$0.9~E_{F}$ (depletion) and $R_{p}$=$150~\mathrm{nm}$; the blue profile corresponds to a reversed perturbation potential (accumulation; $\varphi_{p}^{max}$=$-0.9~E_{F}$). These profiles are extracted from the conductance mapping obtained when $\varphi_{p}$ is swept in the $(x,y)$ plane. They are presented in c) ($\varphi_{p}>0$) and d) ($\varphi_{p}<0$). The vertical dashed lines correspond to the locations of the hard-walls along the scanned line in a).}
\end{figure}

The presence of the polarized conductive AFM tip is numerically modeled using a Lorentzian-shape perturbation potential $\mathit{\varphi_{p}}(x,y)$ - illustrated in Fig. 1a - parametrized by the position of its center ($x_{tip},y_{tip}$), height $\mathit{\varphi_{p}}^{max}$, and width $\mathit{R}_{p}$ which is half the potential FWHM. 
The superposition of $\mathit{\varphi_{p}}(x,y)$ on the modeled disordered potential $\varphi_{d}$, together with the hard-wall boundaries that mark the edges of the nanodevice, define the potential landscape used in the simulations.

In Fig. 3b, the conductance is computed as $\mathit{\varphi_{p}}$ moves along the axis joining the entrance and the exit contacts (dashed line in Figs. 2a and 3a). Remarkably, when the tip position stands nearby the location of the hard-wall (vertical dashed lines), the conductance significantly deviates from that in the absence of perturbation ($\sim 11~\times\frac{2e^{2}}{h}$), e.g. when the tip stands at the center of the device. Beyond fluctuations originating from the presence of the random disorder, the effect is symmetric as positioning the tip near both T-junctions gives the same result. In other words, the Lorentzian potential has a similar effect on conductance when it modifies either the entry or the exit conditions. 
More importantly, this behavior is somewhat counter-intuitive: while a repulsive potential close to both T-junctions actually helps electrons crossing the overall structure (enhanced conductance), an attractive pertubation reduces their ability to pass through the device.

%while a depleting potential close to the T-junction actually helps electrons to flow in and out of the device, an accumulating perturbation refrains electrons from both entering and exiting. 

Looking further in the simulation results, we observe that reversing the sign of $\varphi_{p}$ essentially reverses the change in conductance. Surprisingly, the back-scattering to the leads, due to current focusing on the hard-wall potential of the antidot (described in Figs. 1a and c), is similar in amplitude to the enhanced transmission due to defocusing (Figs. 1a and b). On the other hand, we observe that the symmetry naturally breaks when the tip locates above the leads. In that case, while depleting the lead strongly reduces the conductance, accumulating electrons has naturally a much weaker effect. Finally, when moving the perturbation from the T-junction area towards the center of the device, the effect on $G$ naturally vanishes over a distance corresponding roughly to $R_{p}$ (Fig. 3b).

Beside moving the tip along the device axis, one can also wonder how sensitive $G$-variations are to the perturbation position in the $(x,y)$ plane. This aspect is examined in the $G$ maps plotted in Figs. 3c and d obtained for locally raised or lowered moving potentials, respectively. The main contrast is observed over the T-junctions as well as over the device leads for a depleting potential. In both $x$ and $y$ directions, this contrast fades away over distances comparable to $R_{p}$. When positioning the perturbation potential over the device arms and their vicinities, the $G$ map is decorated with short characteristic length scale fluctuations which are similar to those reported in previous works \cite{martins2007imaging,crook2003imaging,hackens2006imaging}. This weaker amplitude contrast was attributed to the perturbation of resonant states in the local density of states (LDOS) by the moving potential \cite{martins2007imaging,crook2003imaging}, as well as to the  electrostatic Aharonov-Bohm effect \cite{hackens2006imaging}. Note that here the mapping conditions are not suitable for imaging the LDOS because the moving potential is in the strong perturbation and not in the linear regime discussed in Ref. \onlinecite{martins2007imaging}. In this framework we are not using the scanning gate with microscopy purposes in mind.   

It is now time to compare these predictions with experimental results on the sample described above. Figure 4 summarizes the data in a way to ease the comparison with the simulations. We first scanned the biased tip along a line linking the device leads for two opposite sign polarities. Figure 4a shows, like simulations in Fig. 3b, that a depleting (red) potential, near the border of the inner quantum dot, eases electron injection, while an accumulating potential (blue) located at the same place tends to reduce electron transmission through the device. 

%%% figure 4
\begin{figure}[h!]
\centering
\includegraphics[width=8 cm]{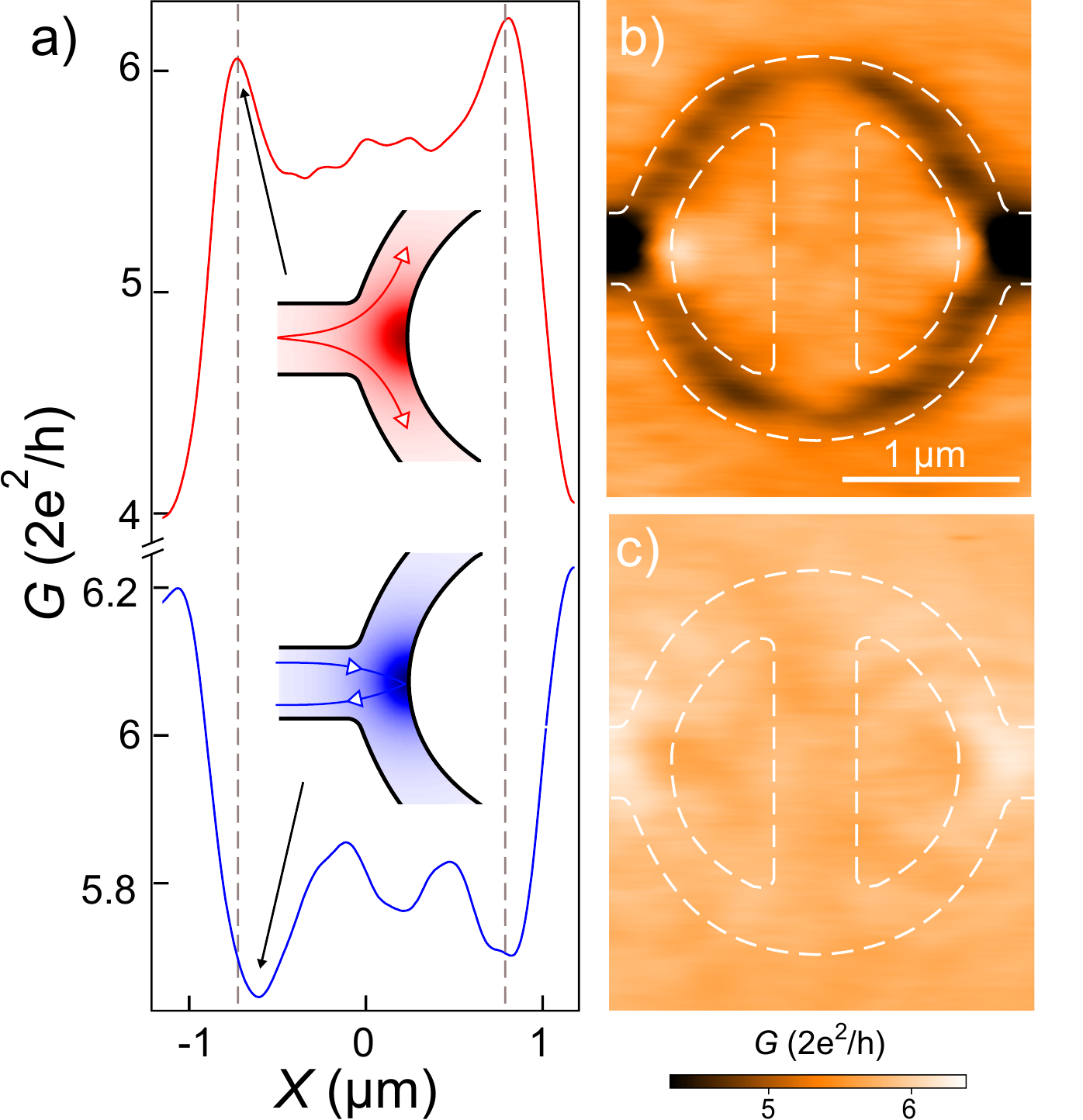} 
\caption{\label{Figure_4}a) Experimental conductance profile as a voltage biased tip is scanned along the dashed line presented in Fig. 2a. The tip is scanned at a distance of 60~nm from the sample surface with $V_{tip}$=$-14~\mathrm{V}$ (red curve) or $+8~\mathrm{V}$ (blue curve). A qualitative scenario is also illustrated for the peculiar electron forth-scattering (red) and back-scattering (blue). b) Conductance map as the polarised tip ($V_{tip}$=$-14~\mathrm{V}$ - depletion) is scanned in a plane at the same constant distance from the sample surface. c) Same map as the one presented in b) but with $V_{tip}=+8~\mathrm{V}$ (accumulation). Note that Fig. S4 presents the same data as c), with an enhanced contrast.}
\end{figure}

For a strongly depleting potential ($V_{tip}=-14~\mathrm{V}$ - red curve in Fig. 4a), corresponding roughly to $\varphi_{p}^{max}\sim 0.4*E_{F}$ (see Fig. S2), $G$ exhibits local maxima when the tip is located above the limit of the etched area in front of the entrance and exit leads (dashed lines in Fig. 4a). As expected, the conductance is reduced all the more as the tip decreases the 2DEG density over the leads. But, counter-intuitively, a strongly accumulating potential ($V_{tip}=8~\mathrm{V}$) brings $G$ to a minimum. Moreover, the effect is essentially symmetric when the tip moves from one T-branch to the other. 
The qualitative match with the curves presented in Fig. 3b (obtained for $\varphi_{p}^{max}=\pm0.9*E_{F}$) is striking, and the experimental conductance maps presented in Figs. 4c and d compare well with the simulations presented in Figs. 3c and d. We observe a remarkable coincidence of simulated and experimental positions and lateral extensions of the peaks and dips located around the hard-walls in the T-junctions.

Resonant features along the ring circumference are also observed in all cases, but the smallest ones that are visible in simulated $G$ maps in Figs. 3c and d, in particular those with concentric shape observed mostly outside the device area, are absent in the experimental data. This is most probably related to thermal averaging, which is not taken into account in the simulations. 

At this stage, we can conclude that the experiments confirm, at least qualitatively, that a focusing/defocusing can be induced by a Lorentzian perturbation combined to a hard-wall potential in a ballistic device. While defocusing (Fig. 4a red) is clearly reminiscent of the Rutherford scattering - here in 2D -, focusing on the hard-wall induces a peculiar back-scattering mechanism as the lensing is combined with the specular reflection illustrated in Fig. 4a (blue). 

At first sight, the weaker absolute value of the voltage applied on the tip in accumulation (blue in Fig. 4a) could explain why the effect on the conductance is weaker than in depletion (red in Fig. 4a). However, we need to dig deeper in the simulations to test the quantitative correspondance between experiments and predictions.

Figure 5 shows the evolution of the conductance when $\varphi_{p}$ travels along the axis of the quantum ring, and when either $\varphi_{p}^{max}$ or $R_{p}$ is varied, the other parameters remaining constant (a similar map with a varying disorder amplitude  $S_{d}$ is shown in supplementary materials - Fig. S1). We obviously focus our attention on the two regions near the edge of the inner QR, i.e. $x\sim\pm$ 800 nm (dashed lines in Figs. 5a and b). We first observe no obvious threshold when $|\varphi_{p}^{max}|$ increases (Fig. 5a). $G$ undergoes a smooth evolution at least up to $2*E_{F}$. However, on the depletion side ($\varphi_{p}>0$), the positions of the local $G$ maxima are gradually shifting towards the center of the device as $\varphi_{p}^{max}$ is made more positive. This reflects the fact that roughly identical potential perturbation conditions are found in the T-junctions both for a weakly perturbing potential ($\varphi_{p}^{max}<E_{F}$) centered close to the hard-wall, and a strongly perturbing potential ($\varphi_{p}^{max}>E_{F})$ centered further away from the hard-wall. On the accumulation side ($\varphi_{p}<0$), the position of the dips' centers remains essentially unaffected : charge accumulation in the T-junctions does not modify its geometry. 

%%% figure 5
\begin{figure}[h!]
\centering
\includegraphics[width=8 cm]{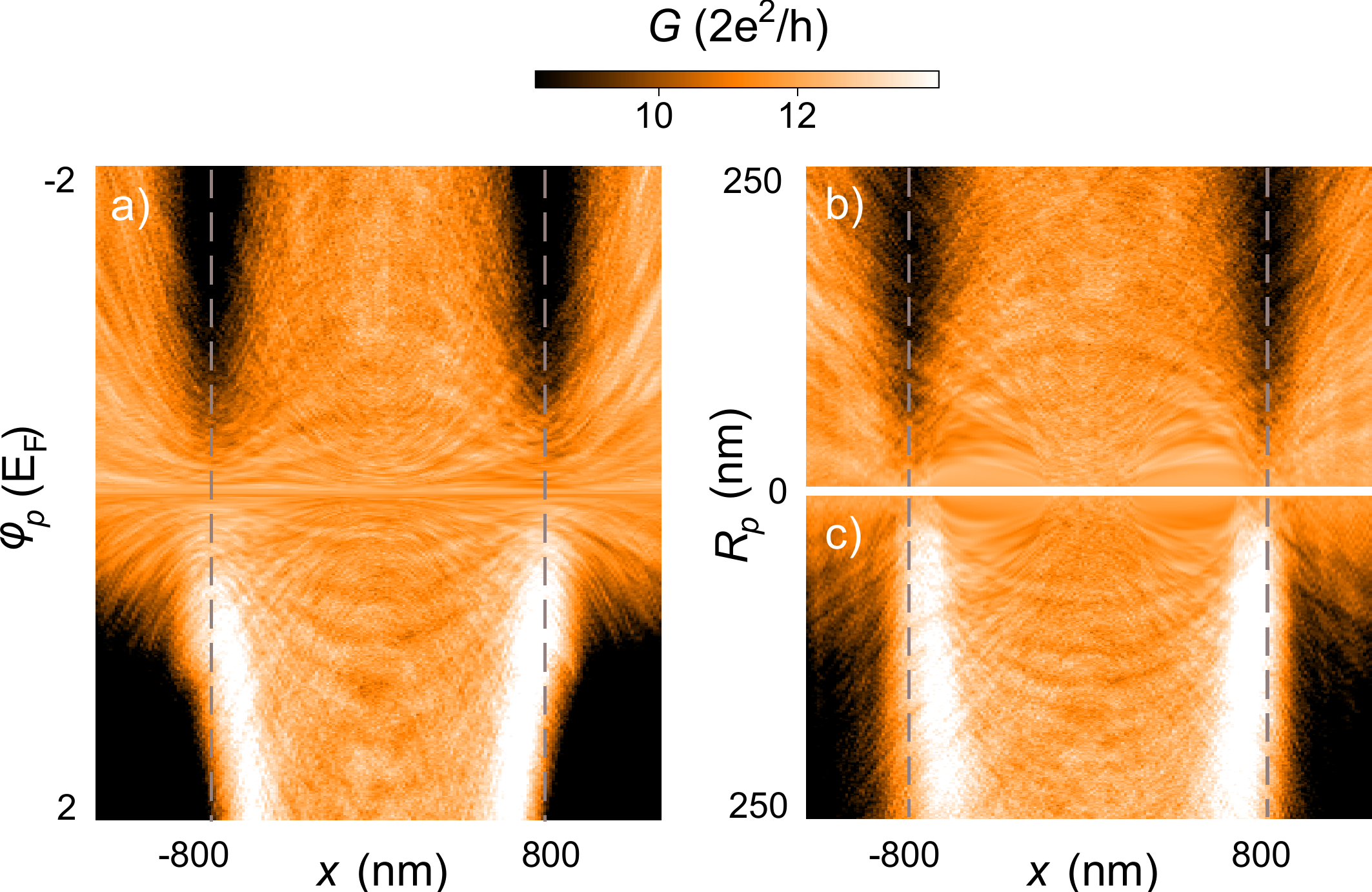}
\caption{\label{Figure_3} Simulated conductance profiles as the potential perturbation is swept along the black dashed line in Fig. 2a for several values of : (a) $\varphi_{p}^{max}$ with $R_{p}$=$150~\mathrm{nm}$ and disorder strength $S_{d}$=$4.78~\mathrm{meV}$, (b) $R_{p}$ with $\varphi_{p}^{max}$=$-0.9~E_{F}$ and $S_{d}$=$4.78~\mathrm{meV}$, (c) $R_{p}$ with $\varphi_{p}^{max}$=$0.9~E_{F}$ and $S_{d}$=$4.78~\mathrm{meV}$. The vertical dashed lines correspond to the locations of the hard-walls along the scanned line.}
\end{figure}

Varying $R_{p}$ has an interesting effect on the conductance peaks and dips. Beyond a few tens of nm, and up to 200 nm where the arms themselves start to be narrowed, varying $R_{p}$ has essentially no effect on the amplitude of conductance extrema, either for negative (Fig. 5b) or positive (Fig. 5c) perturbation potentials. Indeed, the amplitude of conductance peaks and dips saturates for $R_{p}\geq\lambda_F=$ 25 nm, i.e. in the classical regime (Fig. S3).

On the other hand, the evolution of the width of conductance extrema (Figs. 5b and c) is smoother and gives us the possibility to determine the value of $R_{p}^{exp}$ that characterizes our experimental configuration. Based on the FWHM of the strongest (red) conductance peaks in Fig. 4a, we obtain that $R_{p}^{exp}\sim$ 135 $\mathrm{nm}$. This value is well in the range investigated in the simulations and indeed consistent with data discussed in the supplementary informations.

Finally, our results show that increasing the disorder dampens the effect but no qualitative change is observed even when multiplying the initial disorder (Fig. 2b) by a factor of four (see supplementary materials, Fig. S1). This robustness is a clear signature that distinguishes the present effect from universal conductance fluctuations \cite{lee1985universal}, even sensitive to a change of potential amplitude on a single tight-binding site.

To go beyond the good qualitative correspondance between Figs. 3 and 4, we now need to question the experimental data more quantitatively. This is the purpose of Fig. 6 that adresses the effect of the density, or Fermi energy, and finally provides a quantitative comparison between experiments and simulations. 

%%% figure 6
\begin{figure}[h!]
\centering
\includegraphics[width=8 cm]{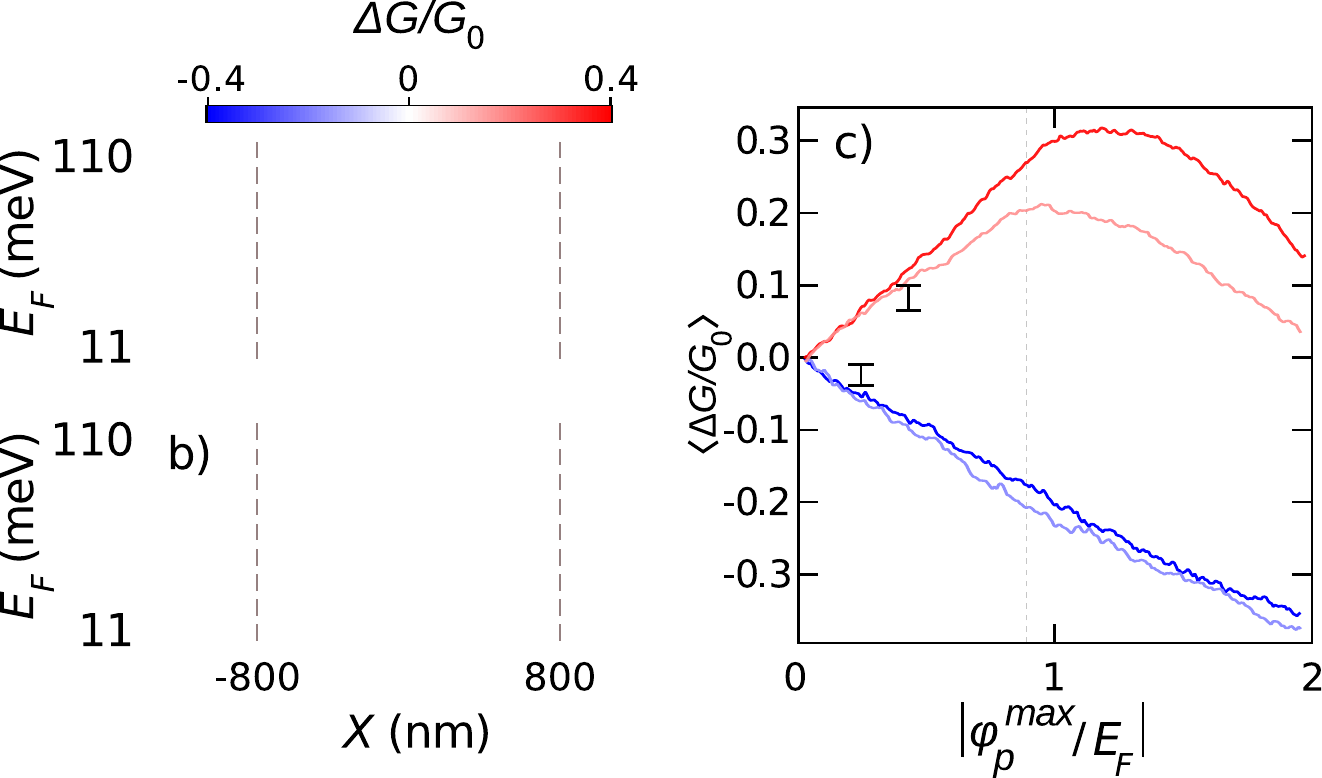} 
\caption{\label{Figure_6}a) and b) Relative variation of conductance ($\Delta G/G_{0}$) with respect to that at $x=0$. c) Relative conductance variation - averaged over the range 110 meV $<E_{F}<$ 11 meV - at $x=800~\mathrm{nm}$ (dark red/blue for depletion/accumulation), and at $x=-800~\mathrm{nm}$ (pale red/blue for depletion/accumulation). The two experimental data points are indicated in c) in the form of two vertical bars.
}
\end{figure}

The variation of the conductance with $E_{F}$, while keeping the absolute value of the ratio $\frac{\varphi_{p}^{max}}{E_{F}}=0.9$, is presented in Figs. 6a and b. Since $G$ increases with $E_{F}$, it makes sense to examine the relative change of conductance $\Delta G/G_{0}=(G-G_{0})/G_{0}$, where $G_0$ is the conductance of the device when the tip is above the device center ($x=0$). It is immediately apparent that $\Delta G/G_0$ is insensitive to $E_{F}$. In other words, the efficiencies of both focusing and defocusing are not sensitive to $E_{F}$ alone, but, as Fig. 6c reveals clearly, to the ratio $\frac{\varphi_{p}^{max}}{E_{F}}$.
More precisely, Fig. 6c shows a linear dependence of $\Delta G/G_0$ as a function of $\frac{\varphi_{p}^{max}}{E_{F}}$ up to $\frac{\varphi_{p}^{max}}{E_{F}}\simeq 1$ in the depletion regime. For $\frac{\varphi_{p}^{max}}{E_{F}}\geq 1$, the symmetry of defocusing with respect to entry and exit breaks down and defocusing becomes less efficient as the arms themselves start to shrink. No such deviation from either linearity or symmetry is observed in the case of depletion (blue lines in Fig. 6c). The counter-intuitive entry/exit symmetry persists in all the range investigated and the linearity with respect to $\frac{\varphi_{p}^{max}}{E_{F}}$ is preserved.

How can we understand this linear dependence, at least in the depletion regime? The defocusing of ballistic electrons facing a Lorentzian-shape repulsive potential is clearly reminiscent of the Rutherford scattering. The original Rutherford formalism provides an expression for the differential cross-section in three dimensions (3D) for a scattering potential $\frac{C}{r}$ - where $C$ is the amplitude and $r$ the distance from the scattering center - as a function of the energy of incident particles $E$ and of the scattering angle $\theta$. Since the arms of the quantum ring capture electrons in a finite angle range from the leads, one can consider the differential cross-section at a given angle as related to the conductance of our ballistic device.
Coincidentally, in the 3D case, the Rutherford formula is independent of wether you treat particles classically or quantumly \cite{friedrich2013scattering}. In 2D, this elegant result is no longer valid in general. In the 2D quantum regime, one has to find an analytical expression of the differential cross-section ($\frac{d\lambda}{d\theta}$) by solving the 2D version of the Lippmann-Schwinger equation \cite{lippmann1950variational} with a Lorentzian-shaped potential distribution, which is far beyond the scope of the present work. 
In the 2D classical regime however, an equivalent formula was derived \cite{barton1983rutherford,friedrich2013scattering}. For the same $\frac{C}{r}$ potential :
\begin{equation}
\frac{d\lambda}{d\theta}=\frac{|C|}{4Esin^{2}(\theta/2)}
\label{equ_1}
\end{equation}
One readily finds from Equ. (1) that, for a given angle $\theta$, the scattering amplitude is fully determined by the ratio between the amplitude $C$ of the perturbative potential and the energy of the particles. In the case of our device, this ratio would correspond to $\frac{\varphi_{p}^{max}}{E_{F}}$. The linear response of $\Delta G/G_0$ with respect to changes in  $\frac{\varphi_{p}^{max}}{E_{F}}$ revealed in Fig. 6c is thus reminiscent of the 2D Rutherford scattering in the classical regime (note that Equ. (1) is also valid in the accumulation regime, but in our QR geometry, specular reflection on the hard-wall must also be taken into account). Beyond Equ. (1) that is probably not strictly applicable to our Lorentzian-shape potential, the Rutherford analogy helps visualizing the observed ballistic defocusing.

We finally turn to what is probably the most important information presented in Fig. 6c : the quantitative comparison between experiments and simulations. To reach that point, we first need to evaluate the amplitude of the perturbation potential induced by the tip. A direct view of the shape of the tip-induced potential experienced by electrons inside the device is obtained by mapping the conductance of a narrow channel in a similar device (whose width is comparable to the leads of the device) close to pinch-off as a function of the electron density, with the tip scanning along a line perpendicular to the channel axis (see supplementary materials, Fig. S2 - this second device is located on the same sample). Following this procedure, we determined that $\varphi_{p}^{max}=3.9~\mathrm{meV}$ for $V_{tip}=-4$ V and $d_{tip}=80~\mathrm{nm}$, and scaled this value taking into account the parameters used in Fig. 3a. Knowing the values of $\varphi_{p}^{max}$ for both the depletion and accumulation potentials data in Fig. 4a, we were able to plot the experimental $\Delta G/G_0$ vs $\frac{\varphi_{p}^{max}}{E_{F}}$ in Fig. 6c.  The good agreement between experiments and simulations reveals the global consistency of our study and that it is indeed possible to strongly enhance or reduce the injection of ballistic electrons in a ballistic device by tuning the shape of the potential faced by electrons. 
It also means that the simple tight-binding model used here captures the essential physics of the phenomena. In the experiment, a conductance change of up to $\sim$10\% relative to the unperturbed device conductance is observed, which is relatively important, compared to e.g. coherent effects at this temperature (40 mK). The phenomenon seems also particularly robust with respect to disorder. This may seem surprising at first sight if its origin is a "ballistic redirecting effect" induced by the tip potential. However, high contrast magnetic focusing effects were observed in semiconductor heterostructures with comparable or lower mobilities \cite{hackens2002long}. This common robustness in both cases further reinforces the idea that ballistic focusing is at the heart of the observed phenomenon.

\section{Conclusion}

In conclusion, we have evidenced surprising ballistic electron focusing and defocusing behaviors governed by a local electrostatic potential. The phenomenology is similar to the 2D Rutherford scattering assuming classical electron dynamics. The applicability of this relatively simple classical formalism in the case of a 2DEG-based device was not expected. Indeed, the scattering amplitude for the interaction between charged particles and a sharp electrostatic potential should in principle be governed by complex interactions related to the presence of the many-particle background of the Fermi sea \cite{Saraga2005}. Other unexpected results of this work resides in two symmetries. The first symmetry concerns the effect of the scattering potential with respect to incoming and outgoing electrons in the T-junctions (i.e. the left-right symmetry in the simulated results). While it is quite straightforward to understand the focusing or defocusing effect of a locally accumulating or depleting potential for incoming electrons, one could not anticipate that a similar effect would be visible for outgoing electrons (i.e. not impinging the hard wall close to normal incidence), in particular in the case of an accumulating potential. A second unexpected symmetry was revealed between the amplitude of the Rutherford defocusing effect (when a depleting potential is applied) and reflective focusing, as experienced by electrons scattered by an accumulating potential in front of a hard wall. All these puzzling fundamental questions will require additional scrutiny and will probably foster further experimental and theoretical work. 

In a broader context, our observations help in the understanding of charge carrier injection in ballistic devices, as it shows that fine tuning of the potential in the vicinity of the entrance and exit leads can have huge effects on transmission through the whole device. In turn, this work provides useful tools in the perspective of building 'electron optics' devices, where a local modulation of the electrostatic potential inside a device redirects the electron flow in a similar way as a optical lens curves light rays \cite{boggild2017two}. In this framework, scanning gate microscopy can play an important role, as pointed out in various theoretical proposals where scattering is investigated by tuning the electrostatic potential at the local scale using a charged metallic tip \cite{Saraga2005,Braun2008,Cserti2007}.
Although the description of scattering in two spatial dimensions was considered as a curiosity up to the early eighties \cite{barton1983rutherford}, nowadays high mobility two-dimentional charge systems gives this fundamental question a complete relevance and the possibility of testing this description, even with relativistic Dirac particles \cite{wu2014scattering,russo2008observation,cabosart2014imaging,cabosart2017recurrent}, also opens new directions of research. 

\section*{Acknowledgements}

This work was funded by the Fonds de la Recherche Scientifique FRS-FNRS (Grants No. J.0067.13, T.0172.13, 326 U.N025.14, J.0009.16, and 2450312F) and by the Communaut\'e Fran\c{c}aise de Belgique (ARC Grant No. 11/16-037, Stresstronics Project and ARC Grant No. 16/21-077, NATURIST Project). S.T. is funded by a Fonds pour la Formation \`a la Recherche dans l'Industrie et dans l'Agriculture FRIA fellowship. B.H. is FRS-FNRS research associate. Computational resources have been provided by the supercomputing facilities of the Universit\'e catholique de Louvain (CISM/UCL) and the Consortium des Equipements de Calcul Intensif en F\'ed\'eration Wallonie Bruxelles (CECI) funded by the Fonds de la Recherche Scientifique de Belgique (F.R.S.-FNRS). S. T. addresses a special thank to D. Fran\c{c}ois for his valuable help concerning parallel computing.

\bibliography{biblio}

\end{document}